\documentclass[aps,prb,groupedaddress,showpacs,twocolumn]{revtex4}

\usepackage{graphicx}
\usepackage{amsmath}
\usepackage{bbm}
\usepackage{xspace}
\usepackage{color}
\usepackage{accents}

\DeclareMathOperator{\Tr}{\mbox{Tr}}
\DeclareMathOperator{\tr}{\mbox{tr}}

\DeclareMathOperator{\gdisc}{{G_{disc}}}
\DeclareMathOperator{\hatgdisc}{{\hat{G}_{disc}}}

\definecolor{orange}{RGB}{252,77,6}
\definecolor{brown}{RGB}{200,127,50}
\newcommand{\noeach}[1]{{#1}}
\newcommand{\nomkch}[1]{{#1}}
\newcommand{\oldwvl}[1]{{ #1}}

\newcommand{\se}{Sec.\@\xspace}

\newcommand{\ie}{i.\thinspace{}e.\@\xspace}

\newcommand{\ptl}{\partial}

\newcommand{\FDF}[2]{\frac{\delta #1}{\delta #2}}

\newcommand{\ve}[1]{{\bf #1}}
\newcommand{\mat}[1]{\mathsf{#1}}
\newcommand{\nag}{{\phantom{\dagger}}}

\newcommand{\eq}[1]{Eq.\thinspace{}(\ref{#1})}

\newcommand{\fig}[1]{Fig.\thinspace{}\ref{#1}}

\newcommand{\fc}[1]{({#1})}
\newcommand{\figc}[2]{Fig.\thinspace{}\ref{#1}\thinspace{}\fc{#2}}

\newcommand{\Figc}[2]{Figure \ref{#1}\thinspace{}\fc{#2}}

\def\ket#1{\mathinner{|{#1}\rangle}}

\newcommand{\Omegas}{\Omega_s}
\newcommand{\tOmegas}{{\widetilde\Omega}_s} 
\newcommand{\hattOmegas}{\hat{\widetilde\Omega}_s} 
\newcommand{\OOmegas}{\Omega_s}
\newcommand{\hatOOmegas}{{\hat\Omega}_s}
\newcommand{\Omegano}{\Omega}
\newcommand{\ho}{{\ve t}}
\newcommand{\unity}{\mathbbm{1}}
\newcommand{\lnm}[1]{\ln\left(-#1\right)}
\newcommand{\lnp}[1]{\ln\left( #1 \right)}
\newcommand{\ginf}{G_{\infty}}
\newcommand{\bdag}[1]{\bar{#1}} 
\newcommand{\hatbdag}[1]{\hat{\bdag{#1}}} 
\newcommand{\cdag}[1]{\bar{#1}} 
\newcommand{\TT}{{\cal T}}

\newcommand{\beq}{\begin{equation}}
\newcommand{\eeq}{\end{equation}}
\newcommand{\beqn}{\begin{eqnarray}}
\newcommand{\eeqn}{\end{eqnarray}}

\newcommand{\eeqref}[1]{Eq.\thinspace{}\eqref{#1}}

\newcommand{\expA}{\mathcal{A}}

\newcommand{\intfuo}[1]{\int \mathcal D #1 }
\newcommand{\de}{\partial}

\usepackage{hyperref}

\newcommand{\calf}{{\cal F}}

\newcommand{\taglia}[1]{}

\begin{document}

\title{Extended self-energy functional approach for strongly-correlated lattice bosons \\in the
  superfluid phase}

\author{Enrico Arrigoni}
\email[]{arrigoni@tugraz.at}
\affiliation{Institute of Theoretical and Computational Physics, Graz University of Technology, 8010 Graz, Austria}
\author{Michael Knap}
\affiliation{Institute of Theoretical and Computational Physics, Graz University of Technology, 8010 Graz, Austria}
\author{Wolfgang von der Linden}
\affiliation{Institute of Theoretical and Computational Physics, Graz University of Technology, 8010 Graz, Austria}

\date{\today}

\begin{abstract}
Among the various numerical techniques to study the physics of
strongly correlated quantum many-body systems, the self-energy
functional approach (SFA) has become increasingly important. In its
previous form, however, SFA is not applicable to Bose-Einstein
condensation or superfluidity. In this paper we show how to overcome
this shortcoming.  To this end we identify an appropriate quantity,
which we term $D$, that represents the correlation correction of the
condensate order parameter, as it does the self-energy for the Green's
function. An appropriate functional is derived, which is stationary at
the exact physical realizations of $D$ and of the self-energy.  Its
derivation is based on a functional-integral representation of the
grand potential followed by an appropriate sequence of Legendre
transformations. The approach is not perturbative and therefore
applicable to a wide range of models with local interactions.  We show
that the variational cluster approach based on the extended
self-energy functional is equivalent to the ``pseudoparticle''
approach introduced in
\href{http://link.aps.org/doi/10.1103/PhysRevB.83.134507}{Phys. Rev. B,
{\bf 83}, 134507 (2011)}. We present results for the superfluid
density in the two-dimensional Bose-Hubbard model, which show a
remarkable agreement with those of Quantum-Monte-Carlo calculations.
\end{abstract}

\pacs{64.70.Tg, 67.85.De, 03.75.Kk}
\maketitle

\section{\label{sec:introduction}Introduction}

Seminal experiments with ultracold gases of atoms trapped in optical
lattices shed new light on strongly-correlated many body
systems.\cite{jaksch_cold_1998,greiner_quantum_2002,bloch_many-body_2008}
In these experiments specific lattice Hamiltonians can be engineered
and investigated with a remarkable high level of control, making
quantum mechanical interference effects observable on a macroscopic
scale. Most important as well as fundamental is the quantum phase
transition of strongly correlated lattice bosons from the localized Mott
phase to the delocalized superfluid phase. In the superfluid phase a
macroscopic fraction of the particles condenses into one quantum
mechanical state, thus, forming a Bose-Einstein condensate, \oldwvl{where the 
number of particles in \nomkch{the} condensate is not necessarily equal to the number of 
superfluid particles.}
In experiments with
ultracold gases of atoms trapped in optical lattices, the condensate
density can be extracted from time-of-flight
images,\cite{greiner_quantum_2002} which are related to the momentum
distribution of the confined particles. Importantly, the finite
expansion time of the particle cloud has to be taken into account when
drawing the connection between these time-of-flight images and the
true momentum
distribution.\cite{gerbier_expansion_2008,trotzky_suppression_2010,kato_sharp_2008,diener_criterion_2007}
However, it is probably even more challenging to measure the
superfluid density itself, as it is not a ground state property but rather
related to the response of the system to a phase twisting
field.\cite{roth_superfluidity_2003} Interestingly, only very recently
for Bose gases without the periodic lattice potential an optical
method has been \nomkch{proposed} 
to extract the superfluid density. \oldwvl{This experiment creates} a vector potential, that imposes angular momentum on normal
fluid particles, while superfluid particles stay at
rest.\cite{cooper_measuring_2010} 

In a previous work, we extended the variational cluster approach
(VCA), which is capable to deal with strongly-correlated many body
systems
\oldwvl{without broken symmetry to the}
superfluid phase of lattice
bosons.\cite{kn.ar.11} Originally, VCA has been formulated
for the normal Mott phase of lattice bosons in
Ref.~\onlinecite{koller_variational_2006} within the so-called
self-energy functional
approach (SFA), which was previously introduced for interacting fermionic
systems.\cite{potthoff_self-energy-functional_2003-1,potthoff_self-energy-functional_2003}
Our extension to the superfluid phase
in Ref.~\onlinecite{kn.ar.11} follows a
different path, and is
based on the so-called ``pseudoparticle'' formalism. 
Within this approach we obtained the
expressions for the superfluid order parameter, the anomalous Green's
function,  and the grand potential, which is the starting point for
the
variational principle, see Eq.\thinspace{}(1), (33), and (2) in
this reference.
 
It should be pointed out that, 
while the pseudoparticle formalism is equivalent to VCA in the normal
phase of both bosonic~\cite{kn.ar.11}
 and
fermionic~\cite{za.ed.02} systems, it lacks the  rigorous 
\oldwvl{theroretical framework}
provided by SFA\nomkch{. In particular,} there is no \nomkch{genuine} variational principle 
\nomkch{explaining} why one should look for a saddle point in the grand potential.
The goal of the present paper is to put th\nomkch{e results obtained within the pseudoparticle approach} \oldwvl{into}  a rigorous framework
by developing an extended \nomkch{self-energy} functional approach\nomkch{, which is capable} to deal with
the bosonic superfluid phase. 

\nomkch{From the present work it will become clear, that} this
extension is not straightforward, as it involves 
{a precise sequence}
 of Legendre
transformations with suitably chosen variables. 
In the search for the appropriate set of transformations the knowledge of the final results provided by pseudoparticle formalism \nomkch{proves} to
be useful.
This fact emphasize\nomkch{s} the advantage of th\nomkch{e} heuristic, yet straightforward, \nomkch{pseudoparticle} approach 
to formulate extensions of VCA.\cite{kn.ar.11} 

The extended SFA formulated in the present paper yields the same
expressions for the superfluid order parameter, for the Nambu Green's
function, and for the grand potential, as obtained from the
pseudoparticle approach. While this might \nomkch{not} seem \nomkch{to be}
surprising, \nomkch{since} we were guided by the 
very results of the
pseudoparticle approach, 
we argue below, that 
our SFA extension presented here is unambiguous.
 The \nomkch{most} important step in this SFA extension
is to find a quantity, which we call $D$, which is the companion of the
self-energy in the superfluid phase. Correspondingly, one has to find
an appropriate universal functional of this quantity and of the
self-energy, which generates the superfluid order parameter and the
Green's function.

As an application, we present an evaluation of 
the superfluid density within this extended VCA, by the 
usual method of introducing a phase twisting \nomkch{field, which}
is equivalent to the helicity
modulus\cite{fisher_helicity_1973} and \nomkch{to} winding
numbers in quantum Monte Carlo (QMC)
algorithms.\cite{pollock_path-integral_1987,prokofev_two_2000}
\nomkch{We evaluate} the superfluid density for the two-dimensional 
Bose-Hubbard (BH) model\cite{fisher_boson_1989}
\nomkch{and find} a \nomkch{remarkably} excellent agreement with QMC results.

This article is organized as follows. In \se~\ref{sec:m1} we extend
SFA to the superfluid phase and obtain the corresponding
extended self-energy functional, along with the appropriate variable
describing superfluidity.
 The evaluation of the
superfluid density within this extended VCA is presented in
\se~\ref{sec:results} and applied to the BH model in two
dimensions. \nomkch{The VCA results} are compared
with unbiased QMC results showing
excellent agreement. Finally, we conclude and summarize our findings in
\se~\ref{sec:conclusion}.

\section{\label{sec:m1}Self-energy functional approach}

Let us recall the key idea of SFA due to M. 
Potthoff\nomkch{.~\cite{potthoff_self-energy-functional_2003-1}}
\nomkch{The starting point} is an appropriate functional 
{
\begin{align}\label{eq:omega}
\hat\Omega[\Sigma, G_{0}^{-1}, H_{U}] &\equiv
\hat{\cal F}[\Sigma, H_{U}] + \hat {\cal E}[\Sigma,G_{0}^{-1}]\;,
\end{align}
}
which consists of a {functional} $\hat {\cal F}$ of the
self-energy, the Legendre transform of the Luttinger-Ward functional,
which is universal 
\noeach{in the sense that 
 it depends on the  interaction part 
($H_U$)}
of the Hamiltonian
but not on the single particle part. The latter enters via the free
Green's function $G_{0}^{-1}$ in the second  functional, which is
explicitly known 
\noeach{
\begin{align*}
\hat {\cal E}[\Sigma,G_{0}^{-1}] &\equiv - \beta \tr \ln(\Sigma - G_{0}^{-1}) 
\end{align*}
}

The functional \nomkch{$\hat\Omega[\Sigma, G_{0}^{-1}, H_{U}]$} has three key features, which are crucial for VCA. 
\begin{itemize}
\item[a)] The non-universal part $\hat{\cal E}$ enters additively in form of a known functional and the many-body aspects are described by a universal functional independent 
of the single particle \nomkch{H}amiltonian, or\noeach{, equivalently, \nomkch{independent} of} $G_{0}^{-1}$. 
\item[b)]
The self-energy of the physical system, characterized by $H_{U}$ and
$G_{0}^{-1}$ is a stationary point of the functional $\hat\Omega$
\noeach{with respect to} $\Sigma$.
\item[c)]
The value of $\hat\Omega$ at the stationary point is 
equal to the thermodynamic grand potential.
\end{itemize}

Given these properties, one can construct a parametric family of \nomkch{H}amilton operators based on the same interaction part (reference systems), for which the thermodynamic grand potential, the
Green's function and the self-energy  can be determined exactly\nomkch{.} 
This allows to determine the exact self-energy functional for self-energies accessible by the reference systems. In this very subspace, the self-energy functional in 
\eq{eq:omega} for the physical system is \noeach{replaced} by that of the reference system. The stationarity condition in turn  allows to determine the Green's function and self-energy of the physical system.

Our goal is to generalize this approach \noeach{to} the superfluid phase as well.
Besides the self-energy, which is the interaction correction of the inverse Green's function, we need the corresponding companion that describes the interaction correction to the order parameter, which we call $D$.

Once the appropriate  form of $D$ has been determined,
we need a functional 
\begin{align*}
\hatOOmegas[\Sigma,D, F, G_{0}^{-1}, H_{U}] &\equiv
\hat {\cal F}[\Sigma,D, H_{U}] + \hat {\cal E}[\Sigma,D,G_{0}^{-1},F]\;,
\end{align*}
in the self-energy $\Sigma$ and  $D$ with the following features. 
\begin{itemize}
\item[a)]  $\hat {\cal F}$ is again a universal functional, now in $\Sigma$ and $D$\nomkch{. T}he non-universal part $\hat {\cal E}$ is explicitly known and carries the dependence on $G_{0}^{-1}$ and the symmetry breaking source-field $F$.
\item[b)]
The functional is again stationary at the exact  
self-energy $\Sigma$ and the exact $D$ of the physical system,
characterized by $H_{U}$\nomkch{,} $G_{0}^{-1}$ and $F$.
\item[c)]
The value of $\hatOOmegas$ at the stationary point
is equal to the thermodynamic grand potential.
\end{itemize}

The sought-for functional $\hatOOmegas$, to be derived in this section, will turn out to be
\noeach{(see below for a definition of the quantities)}
\begin{align}
 2\beta \hatOOmegas[\Sigma,\, D,G_0^{-1},F] &=
 \hat \calf[\Sigma,D] + \hat{\cal E}[\Sigma,\, D,G_0^{-1},F]\\
 \hat{\cal E}[\Sigma,\, D,G_0^{-1},F] &\equiv
 \nomkch{\beta}\Tr\ln [{(\oldwvl{ G_0^{-1}}-\Sigma) \ \ginf}] \nonumber \\
 &+  (\bdag{D} - \bdag{F})( G_0^{-1}-\Sigma)^{-1} (D- F) \;.
\label{oomega}
\end{align}
In the normal phase, it is identical to the \noeach{ functional
  introduced by Potthoff}. The
additional factor 2 is due to the Nambu 
Green's functions. 
\noeach{Moreover, the expression for the grand potential obtained with the help
  of a so-called reference system, see Eq.~\ref{omegafinal}  below, is identical to
  the one obtained within the pseudoparticle approach.~\cite{kn.ar.11}
}
\subsection{Derivation of the grand potential functional}

We start out from the  partition function $Z$ of a 
bosonic many-body system\nomkch{, which} in
a functional integral representation reads 
\begin{equation}
\label{intz}
 Z  = \intfuo{A} \;e^{-S}\;,
\end{equation}
\nomkch{where $S$ is t}he action, \nomkch{which in general} can be written \nomkch{as}\cite{symm}
\begin{align}
 S = &- \frac{1}{2} \int d\tau \int d\tau'
 \,
 \bdag{A}(\tau') \,
 G_0^{-1}(\tau',\tau) 
\, A(\tau) \nonumber \\
&-  \int d\tau \left[
  \bdag{F}(\tau) \, A(\tau) 
-  H_U(A(\tau))\right]\;.
\label{ssy}
 \end{align}
In view of treating the superfluid phase we have adopted a
 Nambu notation \nomkch{in} which the bosonic fields are expressed in a
 vector representation
\beq
 A(\tau) \equiv 
\left( 
 \begin{array}{c}a_1(\tau)\\ \vdots \\
    a_N(\tau) \\ 
\cdag{a}_1(\tau) \\
 \vdots\\  
\cdag{a}_N(\tau) 
\end{array}
\right)\;.
\label{at1}
\eeq
\nomkch{T}he indices $1$ through $N$ denote the corresponding
single-particle orbitals (for example, lattice
sites) \nomkch{where} the boson operators act,
and $a_i(\tau)$ ($\cdag{a}_i(\tau)$) are the fields associated with the annihilation 
(creation) of a boson \nomkch{in} the orbital $i$.
The adjoint field is defined as
\beq
 \bdag{A}(\tau) \equiv \left(
\cdag{a}_1(\tau), \\ \cdots,\\  \cdag{a}_N(\tau),
a_1(\tau),\\ \cdots, \\a_N(\tau)
 \right) \;.
\label{at}
\eeq
\nomkch{It} can be expressed in terms of $A(\tau)$
with the help of the matrix $\TT$\nomkch{,}
 which exchanges the first
$N$ entries of a vector with the last $N$ ones:
\begin{equation}
 \bdag{A}(\tau) = A(\tau)^T \TT \;.
\label{eq:tt}
\end{equation}
\nomkch{The operator} $\TT$ has the properties \nomkch{$\TT^2 = \unity$}, and $\TT = \TT^T$.
The action in \eeqref{ssy} also 
contains \nomkch{the}
 source fields 
\[
\bdag F \equiv \left(f_1,\cdots f_N,\cdag{f}_1,\cdots,\cdag{f}_N\right)
 \quad \text{and} \quad
F=\TT \bdag F^T \;,
\]
\nomkch{which are zero for the physical system of interest,} the boson interaction described by
$H_U$, as well as 
the 
$2N\times 2N$
noninteracting Green's function matrix $G_0(\tau',\tau)$.
\eeqref{intz} with \eeqref{ssy} defines the corresponding grand potential
as a functional of $G_0^{-1}$ and $F$
\beq
\label{somega}
 \hattOmegas[G_0^{-1},F] \equiv
 -\frac{1}{\beta} \ln \oldwvl{\hat Z}\,,
\eeq
where $\beta$ is the inverse temperature. 
Here and in the following, we mark functionals with \nomkch{a} hat
``$\hat{\phantom{\cdots}}$'', and omit their arguments when\nomkch{ever} they are obvious.
\nomkch{T}he noninteracting Green's function 
has the matrix structure (see App.~\ref{contlim})
\beq
\label{g0tau}
{G}_0^{-1}(\tau',\,\tau) =
-\delta(\tau-\tau') \left( \begin{array}{cc} \ptl_\tau + \mat \ho &0 \\ 0 & -\ptl_\tau + \mat
  \ho\,  \end{array} \right) \;,
\eeq
where $\ho$ is the single-particle Hamiltonian matrix.

In the following, 
we carry out a sequence of Legendre transformations starting from
$\hattOmegas$, ultimately
leading to a universal functional 
$\hat \calf[\Sigma,D]$ 
of the \nomkch{self-energy} $\Sigma$ and of a suitable quantity
 $D$ defined in \eeqref{eq:def_D}.
\nomkch{The functional} $\hat \calf$ 
  is the generalization of the self-energy
functional~\cite{potthoff_self-energy-functional_2003-1,potthoff_self-energy-functional_2003,koller_variational_2006} to the superfluid \nomkch{phase, where} a
nonvanishing expectation value \oldwvl{$\expA(\tau) \nomkch{\equiv} \langle A (\tau)\rangle$} of the boson operators $A$ \nomkch{exists}.
\nomkch{The functional}  $\hat \calf$ 
has the properties\nomkch{,} see \eeqref{derupsilon}\nomkch{,} that its functional derivatives \nomkch{with respect to} $\Sigma$ and $D$ yield the
disconnected Green's function, and the expectation value 
$\expA$, respectively.
This procedure 
is inspired by Ref.~\onlinecite{pott.06} and extends that approach to the
treatment of the superfluid phase.

We first determine
 the conjugate variables to  $G_0^{-1}$ and to the source fields $F$. 
The functional derivative of $\hattOmegas$ \nomkch{with respect to}
 the noninteracting Green's function yields\cite{symm}
(see \nomkch{App}.~\ref{matrices})

%
\nomkch{
\begin{align*}
 &2\beta\FDF{\hattOmegas}{{G_{0\,ji}^{-1}}(\tau',\tau)} =  -  \frac{2}{\hat Z} \FDF{}{{G_{0\,ji}^{-1}}(\tau',\tau)} \intfuo{A} \times \\
 &\quad\quad\times \exp\Big\lbrace\frac{1}{2} \int d\tilde \tau \int d\tilde \tau' \bdag{A}_{l}(\tilde \tau) \, G_{0\,ll'}^{-1}(\tilde \tau,\tilde \tau') \, A_{l'}^\nag(\tilde \tau') \\
 &\quad\quad  +  \int d\tilde \tau [ \bdag{F}_l(\tilde \tau) \, A_{l}^\nag(\tilde \tau) -  H_U(\bdag{a},\,a)]\Big\rbrace \\
  &\quad= -  \frac{1}{\hat Z} \intfuo{A} \,{ \bdag{A}_j(\tau')\,A_i^\nag(\tau) \exp\left[-S\right] }\\
  &\quad \equiv \hatgdisc_{,ij}(\tau,\tau') \;.
\end{align*}}
Here $\hatgdisc_{,ij}(\tau,\tau')$
is the disconnected 
interacting time-ordered Green's function.
Along with the definition of the connected
Green's function $\hat G[G_0^{-1},F]$  
we obtain
\begin{subequations}\label{ffa}
\begin{align}
 2\beta \FDF{\hattOmegas[G_{0}^{-1},F]}{G_{0}^{-1}} &= 
\hatgdisc
\equiv \hat
 G -
\hat \expA \hatbdag{\expA}\;.
\intertext{For the functional derivative \nomkch{with respect to} $F$ we obtain similarly}
2\beta \FDF{\hattOmegas[G_0^{-1},F]}{\bdag{F}} &=  
- 2\hat \expA[G_0^{-1},F]\;.
\end{align}
\end{subequations}
\oldwvl{
The two functionals $\hat G[G_{0}^{-1},F]$ and $\hat A[G_{0}^{-1},F]$
defined in \eq{ffa}
\noeach{provide}
the exact Green's function $G$ and order parameter $A$
\noeach{for a
 given noninteracting}   Green's function $G_{0}^{-1}$ and 
\noeach{source field $F$ of the system.}
}
The first step \nomkch{toward the universal functional} consists in a
Legendre transformation replacing the variables $F$ with $\expA$.
To this \nomkch{end}, we invert~\cite{inv} the relation \eeqref{ffa}
  making
$F$  a functional  $\hat F[G_0^{-1},\expA]$ 
\nomkch{and} introduce
\begin{equation}
 \hat \Xi[G_0^{-1},\, \expA] = 2\beta \hattOmegas + 2  \hatbdag{F} \expA \;,
\end{equation}
where, as usually in Legendre transformations, the functional dependence on $F$ has been
eliminated in favor of
 $\expA$  by using the inverse function.
It is straightforward to show that 
the corresponding functional derivatives give
\[
 \FDF{\hat \Xi}{2 \expA} = \hatbdag{F}[G_{0}^{-1},A]\;, \quad
 \FDF{\hat \Xi}{{G_0^{-1}}} = 
\hatgdisc[G_{0}^{-1},A]
\;.
\]
Next\nomkch{,} we modify the functional in the following way
\begin{equation}
 \hat {\widetilde \Xi}[G_0^{-1},\, \expA] = \hat \Xi + \bdag{\expA} G_0^{-1}\expA  \;,
\end{equation}
such that we \nomkch{obtain} the connected Green's function \nomkch{from the}
functional derivative \nomkch{with respect to} $G_0^{-1}$. In total we have
\begin{align}
 \FDF{\hat{\widetilde  \Xi}}{2\expA} &= \hatbdag{F}[G_{0}^{-1},A] + \bdag{\expA}
 G_0^{-1}\;,\\
 \FDF{\hat {\widetilde \Xi}}{{G_0^{-1}}} &= \hatgdisc + \expA \bdag{\expA} = \hat G[G_{0}^{-1},A]\;.
\label{fg0g}
\end{align}
The second step is a
Legendre transformation replacing the variable $G_0^{-1}$ with $G$
\beq
 \hat \Pi[G,\, \expA] = \hat{\widetilde  \Xi} - \nomkch{\beta}\Tr (
   G\,\hat{G}_0^{-1} -\unity )\;,
\eeq
where 
 we have expressed $\hat G_0^{-1}$ as a functional  of 
$G$ and $\expA$, by inverting \eeqref{fg0g}\nomkch{.\cite{inv,symm}} 
\nomkch{We} subtract an ``infinite'' constant \nomkch{$\nomkch{\beta}\Tr \unity$} in order to \nomkch{keep $\hat{\Pi}[G,\, \expA]$} finite.
The functional derivatives of the new functional are
\[
 \FDF{\hat{\Pi}}{2\expA} = \hatbdag{F} + \bdag{\expA} \hat G_0^{-1}\;,
 \quad 
\FDF{\hat {\Pi}}{G} =  - \hat G_0^{-1}\;.
\]
\nomkch{Now, we} modify 
the functional such that we get
the self-energy \nomkch{from the}
 functional derivative 
(see \nomkch{App}.~\ref{trlog})
\begin{equation}
 \hat {\widetilde \Pi}[G,\, \expA] = \hat{  \Pi} + \nomkch{\beta}\Tr\lnp{G/\ginf} \;.
\end{equation}
\oldwvl{
This gives
\begin{subequations}
\begin{align}
 \FDF{\hat{\widetilde\Pi}}{2\expA} &= \hatbdag{F} + \bdag{\expA} \hat
 G_0^{-1}\;, \quad
\FDF{\hat {\widetilde\Pi}}{G} =  \hat \Sigma\\
\hat\Sigma &\equiv   G^{-1} - \hat G_0^{-1}\label{eq:def_sigma}\;.
\end{align}
\end{subequations}
Here we have used the Dyson equation as defining equation for the self-energy}.
Further\nomkch{more}, we carry out a third
Legendre transformation  replacing $G$ with $\Sigma$ in the usual way\nomkch{. Thus we} introduce
\begin{equation}
 \hat P[\Sigma,\, \expA] = \hat{\widetilde  \Pi} + \nomkch{\beta}\Tr\Sigma\,\hat G \;
\end{equation}
with the properties
\[
 \FDF{\hat{P}}{2\expA} = \hatbdag{F} + \bdag{\expA} \hat G_0^{-1}\;,
 \quad 
 \FDF{\hat {P}}{\Sigma} =   \hat G\;.
\]
We modify this functional once more so that its derivative yields a new
function $D$, which will be the companion of the \nomkch{self-energy} in our
extended self-energy approach
\begin{equation}
 \hat {\widetilde  P}[\Sigma,\, \expA] = \hat{P} -
 \bdag{\expA} \Sigma \expA \;. 
\end{equation}
\nomkch{The functional derivatives yield} 
\begin{subequations}
\begin{align}
 \FDF{\hat{\widetilde P}}{2\expA} &= \hatbdag{F} + \bdag{\expA} \hat
 G_0^{-1} - \bdag{\expA} \Sigma = \hatbdag{F} + \bdag{\expA} \hat G^{-1}
 \equiv \hatbdag{D} \;,
\label{eq:def_D}
\\
 \FDF{\hat {\widetilde P}}{\Sigma} &=  \hat G - \expA\bdag{\expA} = 
\oldwvl{\hat G_\text{disc}}
\;.
\end{align}
\end{subequations}

\noeach{
Before proceeding, let us discuss the meaning of the function $D$ 
introduced in \eeqref{eq:def_D}. When extending SFA to the
superfluid phase one is looking for a quantity,
which is related to the condensed order parameter and 
which plays a similar
role as the \nomkch{self-energy}, in that it describes the 
deviation
 between the interacting and non-interacting case. 
Thus, this quantity should
vanish in the noninteracting case
($H_U=0$).
 The reason is that 
SFA will eventually amount to an approximation for $\Sigma$ and $D$,
and we require  this approximation to become exact for $H_U=0$. 
Finally,
 $D$ must obviously vanish in the normal phase.
The
expression 
in \eq{eq:def_D} has precisely these features,
since $\bdag{\expA}_{0}=- \bdag{F}G_0$, which is straightforwardly
determined from the Gaussian integral for $H_U=0$ in \eeqref{ssy}.
Interestingly, the pseudoparticle
approach, presented in Ref.~\onlinecite{kn.ar.11}, and which is
based on an intuitive, yet \nomkch{h}euristic approximation, provides the same 
form of $D$ as given in \eeqref{eq:def_D}. 
}

The final
Legendre transformation 
replacing $\expA$ with $D$
yields the desired  functional of the self-energy and $D$. It represents
the generalization of the self-energy functional ($F[\Sigma]$ of
Refs.~\onlinecite{potthoff_self-energy-functional_2003-1} and \onlinecite{koller_variational_2006})
to the superfluid \nomkch{phase} 
\begin{equation}
 \hat \calf[\Sigma,\, D] = \hat{\widetilde  P} - 2 \bdag{D} \hat \expA  \;
\end{equation}
and has the properties
\beq
 \FDF{\hat{\calf}}{\bdag{D}} =-2 \hat \expA[\Sigma,D]\;, \quad  \FDF{\hat {\calf}}{\Sigma}=  
\hatgdisc[\Sigma,D]
\;.
\label{derupsilon}
\eeq
\oldwvl{
Similarly to \nomkch{$F[\Sigma]$} from Refs.~\onlinecite{potthoff_self-energy-functional_2003-1} and \onlinecite{koller_variational_2006},  $\hat \calf$ is
 (for fixed \noeach{$H_U$})
a
{\it universal} functional of $\Sigma$ and $D$ only, 
from which the disconnected Green's function and the order parameter
are obtained by functional derivative, \nomkch{see}
 \eeqref{derupsilon}. 

Given $\Sigma$ and $D$ we can compute by \eq{derupsilon} the corresponding values for $A$ and $\gdisc$. On the other hand, for a specific 
physical system, uniquely defined  by $G_{0}^{-1}$,  $F$ and $H_{U}$,  
the definitions of the self-energy $\Sigma$\nomkch{,}
\eq{eq:def_sigma}\nomkch{,} and the modified order parameter $D$\nomkch{,}
\eq{eq:def_D}\nomkch{,} provide another set of equations, which uniquely
fix  $\Sigma$ and $D$ via the equations 
\begin{subequations}\label{eq:constr}
 \begin{align} 
 \noeach{\hatgdisc[\Sigma,D]} &\overset{!}{=} (G_0^{-1}-\Sigma)^{-1} + \\ &
(G_0^{-1}-\Sigma)^{-1}(D-F)(\bdag{D} -\bdag{F}) (G_0^{-1}-\Sigma)^{-1} \;,\nonumber
\intertext{and}
\noeach{- 2\bdag{\expA}[\Sigma,D]} &\overset{!}{=} -2(\bdag{D}
-\bdag{F})(G_0^{-1}-\Sigma)^{-1}\;.
\end{align}
\end{subequations}
As for the (original) self-energy functional approach, we seek now a
functional, which becomes
stationary at the exact $\Sigma$ and $D$ for specific $G_{0}^{-1}$ and $F$,
and which consists of the universal functional $\hat\calf$ plus a non-universal 
explicit functional of the form
\begin{align*}
2\beta \hatOOmegas[\Sigma,D,G_{0}^{-1},F,H_{U}] &= \hat \calf[\Sigma,D,H_{U}] +
\hat {\cal E}[\Sigma,D,G_{0}^{-1},F]\;.
\end{align*}
In order to yield the correct stationary point, the functional $\hat {\cal E}$ has to fulfill according to \eq{eq:constr} the  equations
\begin{subequations}
\label{eq:constr_b}
\begin{align}
\FDF{\hat {\cal E}}{\Sigma} =&-(G_0^{-1}-\Sigma)^{-1} \\
&-(G_0^{-1}-\Sigma)^{-1}(D-F)(\bdag{D} -\bdag{F}) (G_0^{-1}-\Sigma)^{-1} \;,\nonumber\\
\FDF{\hat {\cal E}}{D} &= 2(\bdag{D}
-\bdag{F})(G_0^{-1}-\Sigma)^{-1}\;.
\end{align}
\end{subequations}
}
With these ingredient\nomkch{s} we can now express the sought-for functional
$\hatOOmegas$ 
as
\begin{align}
 2\beta \hatOOmegas[\Sigma,\, D,G_0^{-1},F] &=
 \hat \calf[\Sigma,D] + \nomkch{\beta}\Tr\ln [{(\oldwvl{ G_0^{-1}}-\Sigma) \ \ginf}] \nonumber \\
 &+  (\bdag{D} - \bdag{F})( G_0^{-1}-\Sigma)^{-1} (D- F) \;,
\end{align}
which obviously fulfills \eq{eq:constr_b}.
It remains to show that, whenever evaluated at the exact $\Sigma$ \nomkch{and} $D$
the functional
$\hatOOmegas$ corresponds, \nomkch{possibly} apart from a constant, to the 
thermodynamic grand potential $\tOmegas$ of the system.
To this end we add up 
 all the  terms used to construct the functional. At the exact values
 of $\Sigma$ and $D$ we have 
\begin{alignat*}{2}
 2\beta {\hatOOmegas}
\bigl|_{exact}
& = &2&\beta \tOmegas +
2 \bdag{F} \expA + \bdag{\expA} G_0^{-1} \expA - \nomkch{\beta}\Tr \left(G
 G_0^{-1} -\unity\right)\\
&&+& \nomkch{\beta}\Tr \lnp{G/\ginf} 
 + \nomkch{\beta}\Tr\Sigma G - \bdag{\expA} \Sigma \expA- 2 \bdag{D} \expA 
\\
&&-& \nomkch{\beta}\Tr\lnp{G/\ginf} + \bdag{\expA} G^{-1} \expA \\
 &= &2&\beta \tOmegas  - 2\underbrace{(\bdag{D} - \bdag{F})}_{\bdag{\expA} G^{-1}} \expA +2 \bdag{\expA} G^{-1} \expA 
 \\
 &= &2& \beta \tOmegas
\end{alignat*}
We can now proceed as in Refs.~\onlinecite{potthoff_self-energy-functional_2003-1} \nomkch{and} \onlinecite{po.ai.03} \nomkch{and}
construct 
a  reference system, 
which
can be solved \noeach{(almost)} exactly.\cite{cutoff} 
\nomkch{The reference system is} described by \nomkch{a} Hamiltonian
$H'$\nomkch{, which shares} the same interaction
\nomkch{$H_U$} as the physical \nomkch{system}, but \nomkch{consists of} different noninteracting Green's
function $G_0^{\prime}$ and source fields $F'$.
The point is \nomkch{the following: D}ue to the fact that $\calf$ is a universal functional, it cancels
out from the difference between 
$\hatOOmegas$ for the physical and the reference system, 
 {\em with the same values of $\Sigma$ \nomkch{and} $D$}. \nomkch{In particular, this gives}
\begin{multline}
\label{diffomega} 
2 \beta \hatOOmegas[\Sigma,D,G_0^{-1},F]-
2\beta \hatOOmegas[\Sigma,D,G_0^{\prime-1},F']
\\=
\nomkch{\beta}\Tr\lnp{ (G_0^{-1}-\Sigma) \ginf} - 
\nomkch{\beta}\Tr \lnp{( G_0^{\prime-1}-\Sigma)\ginf} 
\\
+ (\bdag{D} -  \bdag{F})( G_0^{-1}-\Sigma)^{-1} (D- F)
\\
-  (\bdag{D} -  \bdag{F}')( G_0^{\prime-1}-\Sigma)^{-1} (D- F') \;,
\end{multline}
\nomkch{which} allows to evaluate the functional $\hatOOmegas$
exactly for the physical system as well, however, in a restricted
subspace of $\Sigma$ and $D$, \oldwvl{representable by the parametric family of reference systems. By construction, the optimal values for $\Sigma$ and $D$ 
for the physical system are those of the reference system for the set of optimal variational parameters.
} 

The variational procedure 
then follows and generalizes Ref.~\onlinecite{potthoff_self-energy-functional_2003-1}: \nomkch{F}irst 
a class of exactly solvable reference systems \nomkch{$\hat H'$} with the same
interaction as the physical system
characterized by a continuum of single-particle parameters $\ho'$
and source fields $F'$ \nomkch{is identified}.
In VCA this class is obtained by  
dividing the original lattice into
disconnected clusters with varying single-particle energies and hopping \nomkch{strengths}.
A larger subspace can be reached by adding bath sites.~\cite{potthoff_self-energy-functional_2003}
\nomkch{T}hen the (connected) Green's function $G'$, 
 the order parameter $\expA'$, and the grand
potential $\Omegas'$ of the reference system \nomkch{is evaluated}.
With the help of Dyson's equation \eeqref{eq:def_sigma} 
the self-energy
$\Sigma'$, and with the help of \eeqref{eq:def_D} 
$D'$ \nomkch{is determined}.
By varying $\ho'$ and $F'$ the subspace of 
self-energies and $D$s is spanned,  which is accessible to the reference system {\oldwvl and to which these objects for the physical system are restricted.}
 Within this subspace the functional
$\hatOOmegas$ can be evaluated exactly for arbitrary
$G_0$ and $F$ of the physical system.
For the relevant case $F=0$ we obtain~\cite{noginf}
from \eeqref{diffomega}
\begin{align}
2 \beta \OOmegas 
&= 2 \beta \Omegas' + 
\nomkch{\beta}\Tr \lnm{( G_0^{-1}-\Sigma')} 
\nonumber \\ & 
 -\nomkch{\beta}\Tr\lnm{( G_0^{\prime-1}-\Sigma')} +
 \bdag{D} ( G_0^{-1}-\Sigma')^{-1} D 
\nonumber \\ & 
\label{omega1}
 -\bdag{\expA}'G^{\prime -1}\expA
\;,
\end{align}
which is now a {\em function} of $\ho'$ and $F'$.
\oldwvl{The infinite physical system can break the symmetry
  spontaneously, while in the reference systems of disconnected finite
  clusters, a non-vanishing order parameter can only be \noeach{achieved} by an additional source field $F'$. This 
explains, why a finite $F'$ is required although $F=0$ in the physical system.}
The SFA approximation consists in 
finding a stationary point of 
$\hatOOmegas$ within this subspace of self-energies \oldwvl{and $D$-s.} 
This corresponds, quite generally,  to 
\nomkch{finding} a stationary point with
respect
to $\ho'$ and $F'$ of \eeqref{omega1}, \ie to the equations
\beq
\label{euler}
\frac{\de \Omegano}{\de \ho'}=0 
\quad\quad
\frac{\de \Omegano}{\de F'}=0  \;.
\eeq
Here, we have replaced $\OOmegas$ with 
 $\Omegano\equiv\OOmegas-\frac12 \tr \ho$ \nomkch{which differs} just 
by a $\ho'$- and $F'$-independent constant \nomkch{and thus }does not change the
saddle-point equations. \nomkch{The quantity} 
$\Omegano$ is the grand potential obtained from the normal-ordered
Hamiltonian
(see \nomkch{App}.~\ref{contlim}).
We also introduce the grand-potential of the normal-ordered reference system
$\Omegano' \equiv \Omegas' -\frac12 \tr \ho'$.
\oldwvl{This term is also present in the pseudoparticle approach,~\cite{kn.ar.11} where its 
origin is easily seen.
}
Moreover, for $\tau$-independent fields and Hamiltonian, the expectation values $\expA(\tau)$ are
$\tau$-independent as well, and the Green's functions depend on the
time difference only. 
In this way, we can rewrite
 \eeqref{omega1} as
\begin{align}
 \Omegano &=  \Omegano' -\frac12 \tr(\ho-\ho') - \frac{1}{2} \Tr \lnm G+\frac{1}{2}\Tr \lnm{G'}  \nonumber \\
  &+ \frac12  \bdag{\expA} G^{-1}(\omega_n=0) \expA -\frac12 \bdag{\expA}' G^{\prime-1}(\omega_n=0) \expA'\;,
\label{omegafinal}
\end{align}
where $G(\omega_n)\equiv \int d \ \tau \ G(\tau,0) e^{ i\tau
  \omega_n}$ 
 is the  Green's function
in Matsubara space. \nomkch{The expression for $\Omega$ given in}
\eeqref{omegafinal} is our main result. 
As 
can \nomkch{be} see\nomkch{n}, this expression
is the same as Eq.~(1) in Ref.~\onlinecite{kn.ar.11}, except for 
a different normalization factor\nomkch{, which is the number of
  clusters $N_c$.}
Notice that $N_c h$ \nomkch{in Ref.~\onlinecite{kn.ar.11}} is equal
to $\ho-\ho'$ \nomkch{in} the present paper.
We thus proved that the result obtained within the pseudoparticle
approach \nomkch{in} Ref.~\onlinecite{kn.ar.11} can be equivalently obtained
within a more rigorous ``generalized'' self-energy \nomkch{functional} approach.
While the pseudoparticle approach is quite intuitive, the present
self-energy approach provides a rigorous variational principle, 
explaining why the grand-potential $\Omegano$ has to be optimized with
respect to the cluster parameters  $\ho'$ and $F'$.
In addition, as in SFA for the normal phase, it suggests
 more general approximations in which bath sites are used to enlarge
 the space of possible \nomkch{self-energies}.\cite{potthoff_self-energy-functional_2003,ba.ha.08,ba.ha.09u}
\begin{figure*}
        \centering
        \includegraphics[width=0.95\textwidth]{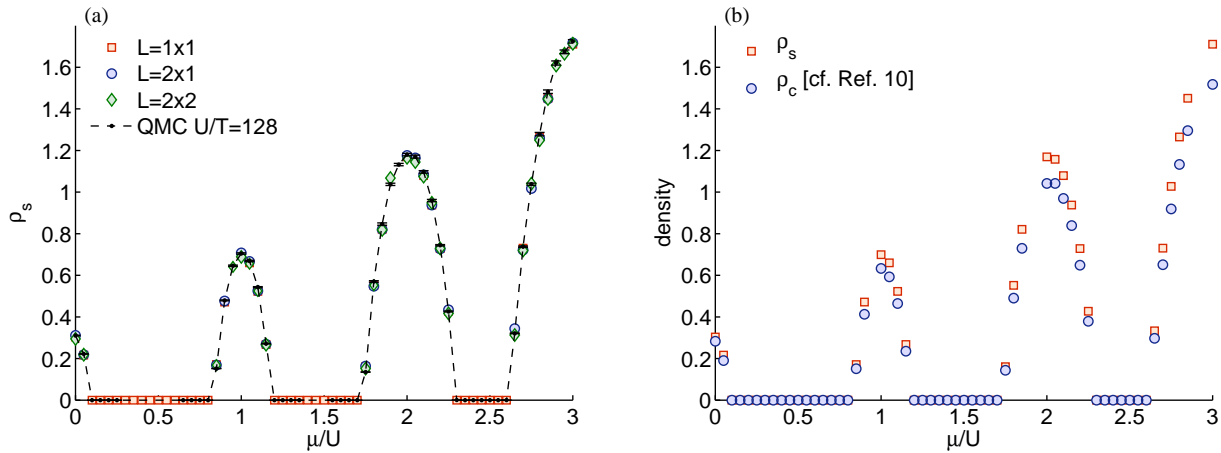}
        \caption{(Color online) Superfluid density $\rho_s$ \fc{a}
          evaluated for constant hopping strength $t/U=0.02$ as a
          function of the chemical potential $\mu/U$. VCA results for
          reference systems of size $L=1\times1$, $L=2\times1$, and
          $L=2\times2$ and for essentially infinitely large physical
          systems are compared to QMC results for physical systems of
          size $32\times32$ and inverse temperature
          $U/T=128$. Comparison of the superfluid density $\rho_s$ and
          condensed density $\rho_c$ \fc{b} for reference systems of
          size $L=1\times1$ and essentially infinitely large physical
          systems, \emph{cf.} Ref.~\onlinecite{kn.ar.11}.
          } 
        \label{fig:stiffnessFixedHopping}
\end{figure*}

\section{\label{sec:results}Superfluid density}

In this section we discuss the evaluation of the superfluid density
$\rho_s$ within our extended \nomkch{SFA/VCA theory} and present results for the
two-dimensional BH model. 

The superfluid density is related to the response of the system to a
phase-twisting
field,\cite{fisher_helicity_1973,lieb_superfluidity_2002} leading to
twisted boundary conditions (BC) in one spatial direction, which we
choose to be the $\ve e_x$-direction, and periodic BC in the
others. The many-body wave function $\ket{\Psi}$ has to obey these BC
and thus 
\[
 \hat T(N_x \, \ve e_x) \ket{\Psi} = e^{i \Theta} \ket{\Psi} \;,
\]
where \nomkch{the operator} $\hat T(\ve r)$ 
\nomkch{translates} the
particles by the vector $\ve r$, $N_x$ is the lattice extension in $\ve
e_x$-direction, and $\Theta$ is the phase twist applied to to the
system. The twisted BC can be mapped
by a unitary transformation onto the lattice Hamiltonian, leading to
complex-valued hopping
integrals.\cite{roth_superfluidity_2003,rey_bogoliubov_2003,poilblanc_twisted_1991}
The resulting Hamiltonian can be interpreted as a cylinder rolled up
along the $x$-direction, which is threaded by an effective
magnetic field with total flux  $\Theta$. 
When a particle is translated by $N_x$ in the  $\ve
e_x$-direction a phase $\exp[-i \Theta]$ is
picked up.\cite{scalapino_insulator_1993} \nomkch{D}ue to gauge
invariance, one is free to choose where the phase is collected when the
particle propagates across the lattice. The usual choice is that each
hopping process in the $\ve e_x$ 
direction, \ie, from site $\ve r' = (r_x-1,\,r_y)$ to $\ve r =
(r_x,\,r_y)$, 
is multiplied by a phase factor
$\exp[-i A]$, where the
associated vector potential is 
\begin{equation}
 A=\Theta/N_x\;.
 \label{eq:vectorpotential}
\end{equation}
\begin{figure}
        \centering
        \includegraphics[width=0.48\textwidth]{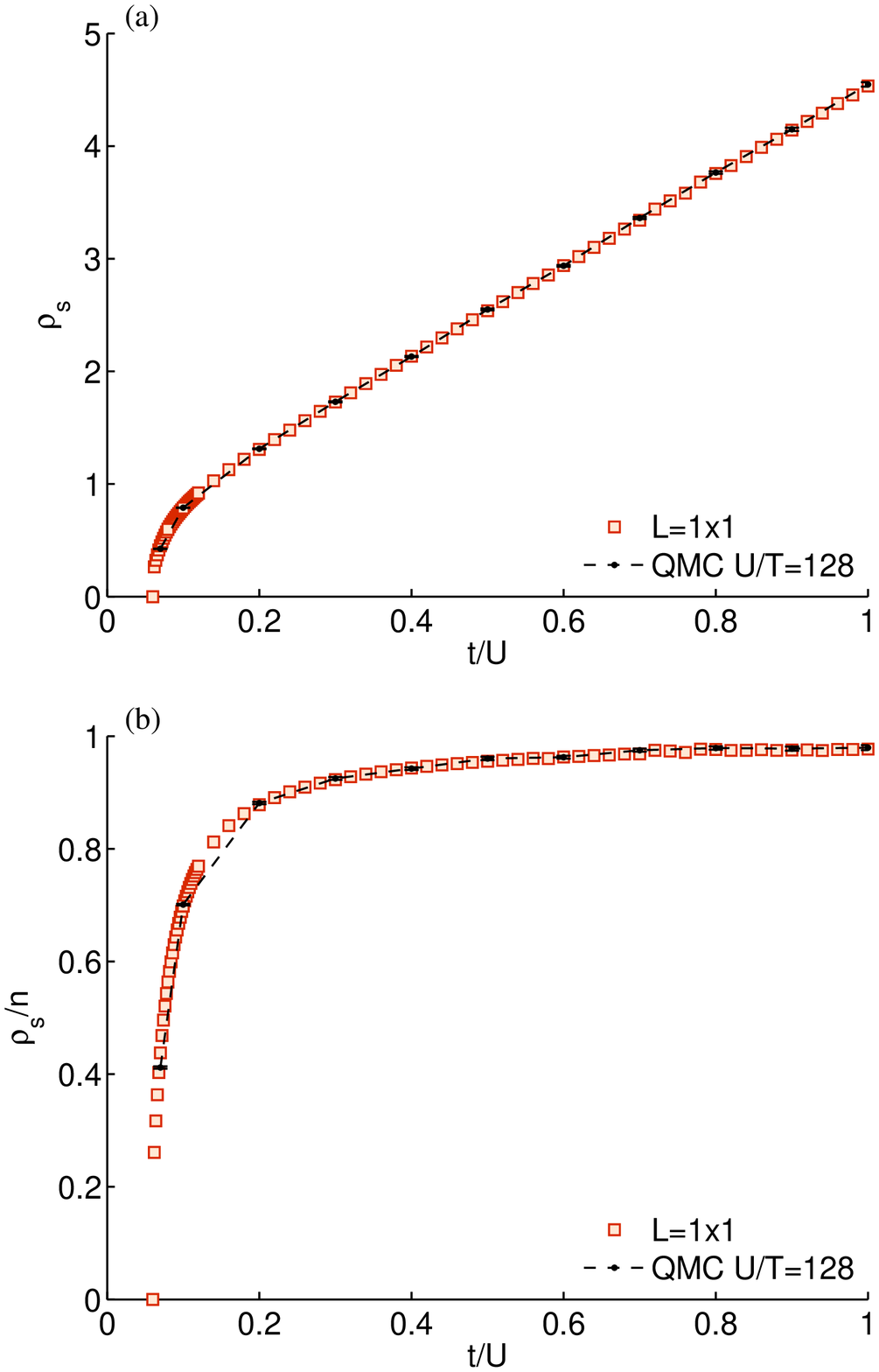}
        \caption{(Color online) Superfluid density $\rho_s$ \fc{a} and
          superfluid fraction $\rho_s/n$ \fc{b} \nomkch{ranging deep in the superfluid phase} evaluated for constant
          chemical potential $\mu/U=0.4$ as a function of the hopping
          strength $t/U$. Results obtained by means of VCA for
          reference systems of size $L=1\times1$ and essentially
          infinitely large physical systems are compared to QMC
          results for physical systems of size $32\times32$ and
          inverse temperature $U/T=128$.  } 
        \label{fig:stiffnessFixedMu} 
\end{figure}
\nomkch{When choosing the phase in that way, the reference system $\hat H'$ also depends on the vector potential $A$ and the intra-cluster hopping terms become complex-valued along the $\ve e_x$-direction. For a} Hamiltonian with nearest-neighbor hopping $t$, 
the superfluid density is determined
from\cite{scalapino_insulator_1993}  
\begin{equation}
 \rho_s = \frac1t \frac{1}{N_x \, N_y} \frac{\ptl^2\, \Omega_{\Theta}}{\ptl\, A^2}\;,
\end{equation}
where $N_x \, N_y$ is the total number of lattice sites of the
physical system, and $\Omega_{\Theta}$ is the grand potential of the
physical system, subject to a phase twist $\Theta$, as discussed
above. 
\nomkch{Plugging in the vector potential of \eq{eq:vectorpotential} yields}
\begin{equation}
 \rho_s = \frac1t \frac{N_{x} }{ N_y} \frac{\ptl^2\, \Omega_{\Theta}}{\ptl\, \Theta^2}\,.
\label{eq:stiffness}
\end{equation}
In practice, the grand potential $\Omega_\Theta$ is evaluated at the
stationary point of \eq{omegafinal}, and is determined
self-consistently for several values of $\Theta$. From this data
the curvature of $\Omega_\Theta$ \nomkch{with respect to} $\Theta$ is extracted from a
fit. Using the curvature, the superfluid density is evaluated
according to \eq{eq:stiffness}. Note that a finite cluster is embedded
in an \oldwvl{essentially} infinitely large system and thus the limits are taken
in the correct order to obtain the superfluid
density.\cite{scalapino_insulator_1993} 

In the following, we apply this procedure to the two-dimensional BH model\cite{fisher_boson_1989}
\nomkch{\[ \hat{H}= \sum_{\left\langle i,\,j \right\rangle} t_{ij}\,a_i^\dag \, a_j^\nag
+ \frac{U}{2} \sum_i \hat{n}_i\left(\hat{n}_i-1 \right) - \mu \sum_i \hat{n}_i \; \mbox{,}
\]}
where $a_i^\dag$ ($a_i^\nag$) creates (destroys) a bosonic particle on site $i$,
and $\hat{n}_i = a_i^\dag\,a_i^\nag$ is the occupation number operator. 
\nomkch{The hopping integrals $t_{ij}$ are nonzero for nearest neighbors only, as indicated by the the angle brackets. Specifically, $t_{ij}=-t$ for hopping processes along the $\ve e_y$-direction and $t_{ij}=-t \exp[i\,A\,(\ve r_i - \ve r_j)\ve e_x]$ for hopping processes along the $\ve e_x$-direction.  }
The chemical potential, termed $\mu$, controls the particle number and
$U$ is the repulsive on-site interaction, which subsequently will be
used as unit of energy. The reference system $\hat H'$ consists of a
cluster decomposition of the physical system $\hat H$ plus a $U(1)$
symmetry breaking source term\nomkch{
\begin{align*}
 \hat{H}'= \sum_{\ve R} \Big[ & \sum_{\left\langle \alpha,\,\beta \right\rangle} t'_{\alpha\beta}a_{\alpha, \ve R}^\dag \, a_{\beta, \ve R}^\nag
+ \frac{U}{2} \sum_\alpha \hat{n}_{\alpha, \ve R}\left(\hat{n}_{\alpha, \ve R}-1 \right) \\
  &- \mu' \sum_\alpha \hat{n}_{\alpha, \ve R} -\sum_\alpha ( a_{\alpha, \ve R}^\dag \, f_{\alpha}^\nag + f_{\alpha}^* \, a_{\alpha, \ve R}^\nag)\Big]\; \mbox{,}
\end{align*}
where the lattice site indices $i$ have been decomposed into an index $\ve R$, that specifies the cluster and into an index $\alpha$, that specifies the lattice sites within a cluster.\nomkch{\cite{kn.ar.11,knap_spectral_2010}} Analogously to the physical system, the hoping integrals are $t'_{\alpha\beta}=-t'$ and $t_{\alpha\beta}'=-t' \exp[i\,A\,(\ve r_{\ve R \alpha} - \ve r_{\ve R \beta})\ve e_x]$ for nearest-neighbor hopping processes along the $\ve e_y$- and the $\ve e_x$-direction, respectively, and zero otherwise. 
In our calculation, we use the chemical potential $\mu'$ and 
the
source coupling strength $f_\alpha$ of the reference system as
variational parameters in the optimization prescription.
Since the reference system is complex valued, the source coupling strength $f_\alpha$ is complex valued too, \ie, $f_\alpha=|f_\alpha| \exp[\phi_\alpha]$. Thus, in general, $2L$ variational parameters have to be considered, where $L$ is the number of cluster sites. However, for different cluster sites $\alpha$ the source coupling strengths $f_\alpha$ are interrelated, as can be seen from mean field arguments, leading effectively to two variational parameters $|f|$ and $\phi$, which we use---in addition to the chemical potential $\mu'$---to treat complex valued reference systems.}

In \fig{fig:stiffnessFixedHopping} we present
the superfluid density $\rho_s$ for different sizes of the reference
system ranging from $L=1\times1$, over $L=2\times1$, to $L=2\times2$
and essentially infinitely large physical systems. 
\nomkch{For the largest cluster we restrict the variational search space
to real valued order parameters, \ie, we set $\phi_\alpha=0$.
\Figc{fig:stiffnessFixedHopping}{a} demonstrates that this choice leads
to comparable results as obtained with the full variational space. Yet,
for the restricted variational space the computational effort as well as
the numerical complexity is reduced, since the reference system remains
real valued.} 
\Figc{fig:stiffnessFixedHopping}{a} shows the superfluid
density $\rho_s$, as a function of the chemical potential $\mu/U$
evaluated for fixed hopping strength $t/U=0.02$.  
The chemical potential ranges from $\mu/U=0$ to $\mu/U=3$. As the
hopping strength is small, three regions with $\rho_s=0$ are present,
corresponding to the Mott insulating phase. In between these regions,
we observe a finite superfluid density $\rho_s$ indicating the
occurrence of the
superfluid phase.  
In addition to the VCA results, we show QMC results with errorbars
(barely visible) for physical systems of size $32\times32$ and inverse
temperature $U/T=128$. The QMC calculations were performed with the
ALPS library\cite{albuquerque_alps_2007} and the ALPS
applications.\cite{ALPS_DIRLOOP} Particularly, we use the stochastic
series expansion representation of the partition function with
directed loop updates,\cite{sandvik_1991,evertz_1993,Sylju_2002} where
the superfluid density is evaluated via the winding
number.\cite{pollock_path-integral_1987,prokofev_two_2000} The
superfluid density $\rho_s$ obtained from VCA agrees remarkable well
with the QMC results. Furthermore, VCA results are almost independent
of the size $L$ of the reference system, signaling convergence to the
correct results even for $L=1\times1$ site clusters. The superfluid
density $\rho_s$ is compared to the condensate density $\rho_c=\langle
a_i \rangle$ in \figc{fig:stiffnessFixedHopping}{b}, \emph{cf.}
Ref.~\onlinecite{kn.ar.11}. It can be observed that the
superfluid density is always larger than the density of the
Bose-Einstein condensate. However, the difference between the two
densities is rather small, since a very dilute Bose gas is
investigated. 

In \fig{fig:stiffnessFixedMu} we evaluate \fc{a} the superfluid density $\rho_s$ and \fc{b} the superfluid fraction $\rho_s/n$ ($n$ is the particle density) for fixed chemical potential $\mu/U=0.4$ as a function of the hopping strength $t/U$. 
The hopping strength ranges from $t/U=0$ to $t/U=1$\nomkch{, which is already very deep in the superfluid phase}. 
For $\mu/U=0.4$ the phase boundary between the Mott and the superfluid
phase is located at $t/U\approx0.06$. In the superfluid phase close to
the phase boundary the superfluid density rises quickly from zero
developing an almost linear behavior for $t/U \gtrsim 0.2$. In the
latter parameter regime the superfluid fraction is larger than $90\%$
signaling that already a very large amount of the lattice bosons is
superfluid. 
\nomkch{As emphasized in
  Ref.~\onlinecite{rancon_strongly-correlated_2010}, 
a relatively sharp
crossover from a strongly-correlated superfluid, characterized by a
superfluid fraction which is well below $1$, to a weakly-correlated
superfluid, where the superfluid fraction is almost $1$, can be
observed, see \figc{fig:stiffnessFixedMu}{b}.} 
 In addition to the
VCA results evaluated for reference systems of size $L=1\times1$ and
essentially infinitely large physical systems, we show QMC results for
physical systems of size $32\times32$ and inverse temperature
$U/T=128$, which again exhibit perfect agreement. 

\begin{figure*}
        \centering
        \includegraphics[width=0.98\textwidth]{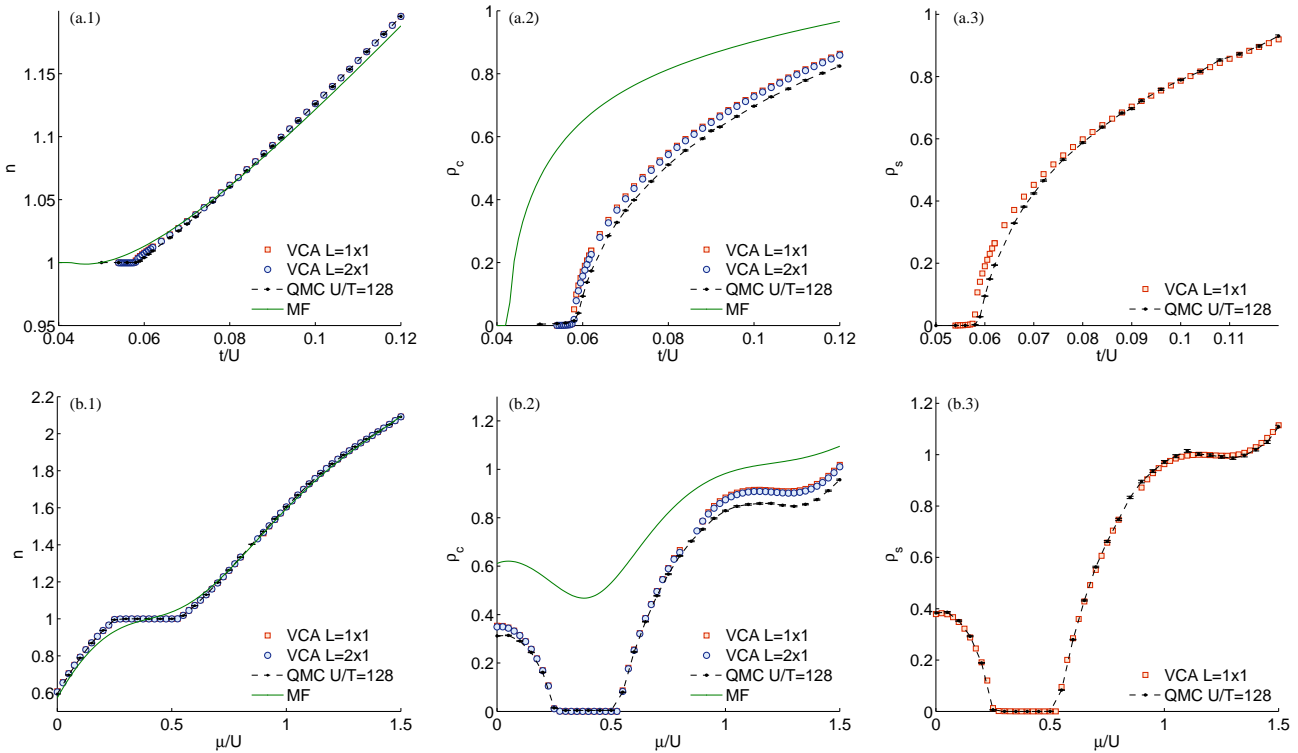}
        \caption{(Color online) \nomkch{Particle density $n$ (left),
            condensate density $\rho_c$ (middle), and superfluid
            density $\rho_s$ (right) evaluated 
\noeach{
around the quantum critical 
}
            region close to the tip of the first Mott lobe. Comparison
            of the data obtained by means of VCA (for essentially
            infinitely large physical systems and reference systems as
            stated in the legends), QMC (for physical systems of size
            $32\times32$ and inverse temperatures $U/T=128$), and
            mean-field. The first row \fc{a.\textasteriskcentered}
            shows results for fixed chemical potential $\mu/U=0.4$ as
            a function of the hopping strength $t/U$, whereas the
            second row \fc{b.\textasteriskcentered} shows results for
            fixed hopping strength $t/U=0.05$ as a function of the
            chemical potential $\mu/U$.}  }  
        \label{fig:closeMott} 
\end{figure*}
\nomkch{In \fig{fig:closeMott} we focus on the quantum critical region
  close to the tip of the first Mott lobe,
\noeach{
 which is the most challenging one. 
}
 In particular, we evaluate the particle density
  $n$, the condenstate density $\rho_c$, and superfluid density
  $\rho_s$. In the first row we show results for fixed chemical
  potential $\mu/U=0.4$ as a function of the hopping strength $t/U$,
  whereas in the second row we keep hopping strength fixed at
  $t/U=0.05$ and vary the chemical potential $\mu/U$. We compare VCA
  results with QMC and mean-field (MF). The most important observation
  is that MF is far off QMC and VCA. For $\mu/U=0.4$ MF predicts the
  phase transition to be at a much smaller value of $t/U$ than QMC and
  VCA. This leads to significant deviations in both the density and
  condensate density as compared to QMC and VCA. For fixed $t/U=0.05$
  MF does not enter the Mott region and thus does not predict a
  plateau in the density. For both investigated situations (fixed
  $\mu/U$ and fixed $t/U$) the results obtained by means of VCA and
  QMC agree quite well. For the QMC simulations we used lattices of
  size $32\times 32$ and inverse temperatures of $U/T=128$. The VCA
  results are obtained at zero temperature for clusters of size
  $1\times1$ and $2\times1$, respectively, and essentially infinitely
  large physical systems. In this challenging regime small differences
  between VCA and QMC are observable for the condensate density. For
  the reference system sizes considered here, results are almost
  identical. 
\noeach{
Larger reference systems might still reduce 
}
  the difference between VCA and QMC. However, close to the phase
  transition finite size and finite temperature effects might still be
  important for the QMC results, and thus a proper 
\noeach{
finite size scaling of these data 
}
  might also reduce the discrepancy between the two approaches. Note
  that for fixed hopping $t/U=0.05$ there is a very small region at
  $\mu/U\approx 0.85$, where it is difficult to numerically determine
  the stationary point of the grand potential. Such a region is also
  present between the first and the second and between the second and
  the third Mott lobe in \fig{fig:stiffnessFixedHopping}. However,
  there it is barely visible since 
\noeach{
the spacing   between two consecutive  $\mu$  datapoints  is larger than this gap. 
}
\noeach{
 This failure 
appears to be related to the fact that two solutions adiabatically
connected to two sectors with different particle numbers, i. e. the
two neighboring Mott regions, meet and try to avoid each other.
}
However, we want to emphasize that this
\noeach{
problem  affects only a tiny region of the phase diagram.  
}
When keeping the chemical potential fixed at
  $\mu/U=0.4$ solutions can be easily found for all values of the
  hopping strength. } 

Finally, we want to emphasize that the VCA results are obtained with
very modest computational effort and that excellent agreement with QMC
can be observed, even for very small reference systems.   

\section{\label{sec:conclusion}Conclusions}
\nomkch{

In the present work, we extend the self-energy functional
approach to the $U(1)$ symmetry broken, superfluid phase of correlated
lattice
bosons. A crucial point of this extension is the identification of a quantity,
termed $D$, which is the companion of the self-energy $\Sigma$ in the
superfluid phase. We also identify the appropriate (nonuniversal) functional
$\hatOOmegas$ 
which is stationary at the physical values of
 the self-energy $\Sigma$ and of  $D$.
In analogy to the self-energy, which is the difference of the interacting
and non-interacting Green's function, the quantity $D$ 
is related to
the difference of the order parameter of the interacting and
non-interacting systems. Thus, $D$ is zero in the normal phase and for
$U=0$. From
these relations also follows that both $\Sigma$ as well as $D$ vanish
in the non-interacting case. Importantly, when the functional
$\hatOOmegas$ is evaluated at the exact values of $\Sigma$ and $D$ it
corresponds to the grand potential of the 
physical Hamiltonian. To evaluate the functional, we proceed as in the
original \nomkch{self-energy} functional approach,\nomkch{\cite{potthoff_self-energy-functional_2003-1}}  and introduce a reference
system, which is a cluster decomposition of the physical
system. Importantly, the reference system shares its two-particle
interaction with the physical system, and can be exactly solved by
numerical methods. By comparison of the functionals,
the universal part of $\hatOOmegas$, denoted as $\hat \calf$, can be
eliminated, which allows to evaluate $\hatOOmegas$ exactly on the
subspace of $\Sigma$ and $D$, spanned by the possible sets of reference systems. 
The results presented are shown to be equivalent to the ones obtained
by a more heuristic method,  the
pseudoparticle approach introduced in
Ref.~\onlinecite{kn.ar.11}, and thus provide rigorous
variational grounds for that approach.
In addition,  the extended
self-energy functional approach introduced here
allows to envision more general
reference systems, in which bath sites are incorporated to enlarge the
space of possible self-energies $\Sigma$, and possibly 
bridge over to (Cluster) Dynamical Mean Field
Theory (DMFT).\nomkch{\cite{potthoff_self-energy-functional_2003-1,ge.ko.96}}
\nomkch{For future research it} would be interesting to verify whether in the limit of an infinite
number of bath sites and for a single correlated site as a reference system,
our superfluid SFA becomes equivalent to DMFT for superfluid
bosons,\nomkch{\cite{by.vo.08,an.gu.10}}
as it is the
case in the normal phase.\nomkch{\cite{potthoff_self-energy-functional_2003-1}}
For a finite number of bath sites this is certainly not the case,
since the order parameter in the reference system differs from the
physical one.

We also presented 
 how the superfluid density can be evaluated by means of
this extended variational cluster approach. To this end we applied a
phase twisting field to the system. We evaluated the superfluid
density for the two-dimensional Bose-Hubbard model and compared the
extended variational cluster approach results with unbiased quantum
Monte Carlo results, yielding remarkable agreement. 
\nomkch{We want to emphasize that the extended self-energy functional
  approach is not only applicable to the Bose-Hubbard model but to a
  large class of 
\noeach{
lattice models, which exhibit a condensed phase. 
}
\noeach{
 This includes experimentally interesting systems such as disordered bosons, 
}
   multicomponent systems (Bose-Bose mixtures or Bose-Fermi
  mixtures) and light matter
  systems.\cite{hartmann_quantum_2008,tomadin_many-body_2010}
\noeach{
Strictly speaking, the method cannot treat long-range interactions\nomkch{, such as dipolar ones,} exactly
.\cite{barnett_quantum_2006,micheli_toolbox_2006} However, the
long-range part can be incorporated on a mean-field level.\cite{aichhorn_charge_2004} 
In principle,  the present approach 
 can be applied to systems with
  broken translational invariance as well, and, for example, \nomkch{can} consider
  the effect of a confining magnetic trap.
However, in this case 
 one has to abandon the
  Fourier transform in the cluster vectors and work in real space and,
  thus, deal 
 with larger matrices and a larger number of variational parameters. 
A convenient, numerically less expensive alternative, is to adopt the so-called
local density approximation.~\cite{kollath_spatial_2004}
}
 } 
}

\begin{acknowledgments}
We made use of the ALPS library and the ALPS applications.\cite{albuquerque_alps_2007, ALPS_DIRLOOP}
We acknowledge financial support from the Austrian Science Fund (FWF)
under the doctoral program ``Numerical Simulations in Technical
Sciences'' Grant No. W1208-N18 (M.K.) and under Project No. P18551-N16
(E.A.).
\end{acknowledgments}

\appendix

\section{Notation and conventions}

\subsection{Matrix notation}
\label{matrices}

\subsubsection{General}

In order to 
simplify our notation 
we  omit time arguments, whenever this does not cause ambiguities.
Therefore, 
two-point functions such as
Green's functions, self-energies, etc. are interpreted as
matrices in Nambu, orbital, 
 and $\tau$ space. 
\nomkch{O}ne-point objects such \nomkch{as} $\expA$ ($\bdag{\expA}$)
are interpreted as 
column (row) vectors in the same space. 
Matrix-matrix and vector-matrix product\nomkch{s} 
are understood throughout, whereby internal $\tau$ variables are \nomkch{considered to be} integrated over.
In addition, the transposing operator ``${}^T$'' also acts on time
variables\nomkch{. T}races $\Tr$ contain an integral over $\tau$
\nomkch{and a trace $\tr$ over orbital indices, \ie,  
$\Tr M \equiv \tr \int_0^{\beta} d \tau M(\tau,\tau+0^+)$}, 
where the $0^+$ leads to the well known 
convergence factor \nomkch{$e^{i \omega_n 0^+}$} in Matsubara space.

(Functional) derivatives \nomkch{with respect to} matrices are defined ``transposed'':
\[\left(\FDF{\hat X}{M}\right)_{ij}(\tau,\tau') \equiv
\FDF{\hat X}{M_{ji}(\tau',\tau)}\;.\]

Finally, there are two types of products between row (in the form $\bdag{v}$)
and column ($u$) vectors, depending on the order:
\nomkch{On the one hand t}he product 
$ \bdag{v} u$
produces a scalar (all indices are summed/integrated over).
On the other hand, inverting the order, as in 
$u \bdag{v}$ gives a matrix, as
indices are
``external''
and, thus, not summed over.

\subsubsection{\nomkch{Trace} in
  $\tau$ \nomkch{and in} Matsubara space}
\label{fourier}
In $\tau$ space we have
\[
\Tr M = \beta^{-1}
\nomkch{\tr} \int_{0}^{\beta} d\tau\ M(\tau,\tau+0^+) \;.
\]
The transformation \nomkch{of $M(\tau,\tau')$} to Matsubara space is defined as
\[
M(\tau,\tau') \equiv
\beta^{-1} \sum_{n,n'} M(\omega_n,\omega_n') e^{-i
  \omega_n \tau+ i \omega_n' \tau'} \;.
\]
The inverse transformation reads
\[
 M(\omega_n,\omega_n') = 
\beta^{-1} \int d \tau d \tau'
M(\tau,\tau')   e^{i
  \omega_n \tau- i \omega_n' \tau'} \;.
\]
\nomkch{Combining the equations above}, the trace becomes 
\begin{align*}
\Tr M &= \nomkch{\tr} 
\int_{0}^{\beta} d\tau \beta^{-2} \sum_{n,n'} M(\omega_n,\omega_n') 
e^{-i  (\omega_n-\omega_n') \tau + i \omega_n' 0^+}
\\ &=
\beta^{-1}
 \sum_n \nomkch{\tr} M(\omega_n,\omega_n)  e^{i \omega_n 0^+}
\;.
\end{align*}

\subsubsection{Logarithm}
\label{trlog}
There are some subtle points concerning 
logarithms of
two-point functions.
Although  these issues are immaterial  for the final result, we prefer
to specify them in detail.

The  logarithm of $G$ considered as a matrix 
in the continuum variable $\tau$  is 
defined up to an infinite constant which depends on the the
discretization step $\delta$ (see below). 
In addition, the trace of the logarithm carried out in Matsubara space
diverges as well (despite the 
convergence factor $e^{i \omega_n 0^+}$). 
The usual result presented in the literature 
(see, \nomkch{for instance} Ref.~\onlinecite{lu.wa.60}) implicitly assumes that an 
infinite constant has been subtracted.
In order to avoid these undetermined infinite terms, we subtract them explicitly
at the outset
with the help of the ``infinite energy'' Green's function 
\begin{align*}
\ginf(\tau,\tau') &= \beta^{-1} \sum_{n} \ginf(\omega_n) e^{-i
  \omega_n(\tau-\tau')}\\
\ginf(\omega_n) &= \unity \frac{1}{i \omega_n -E} \;,
\end{align*}
where it is understood that the $E\to+\infty$ limit is 
taken at the end of the calculation. 
This choice guarantees, for example, that $\Tr \ln G/\ginf$, where \nomkch{$G$} 
is the Green's function in normal (i.e. not Nambu) notation, vanishes in the \nomkch{limit} 
$\mu\to-\infty$, where $\mu$ is the 
chemical potential.

The Fourier transform defined in \nomkch{App}.~\ref{fourier} allows to define
the logarithm of $G$ in $\tau$ space, apart from an
infinite multiplicative constant, which originates from the fact that the
Fourier transformation is not and cannot be normalized in the
continuum limit. \nomkch{In particular,}
\begin{align*}
[\ln (-G)](\tau,\tau') &=
\beta^{-1} \sum_{n,n'} [\ln(-G)] (\omega_n,\omega_n') e^{-i
  \omega_n \tau+ i \omega_n' \tau'} \\
&= \beta^{-1}
 \sum_{n} \ln \left[-G(\omega_n)\right] e^{-i
  \omega_n (\tau-\tau')} \;,
\end{align*}

\subsection{Symmetry of Green's \nomkch{functions} and other two-point functions}
\label{symm}

The action in \eq{ssy} 
is invariant under the transformation
$G_0 \to (\TT  G_0^T \TT)$, where 
the transposing operator ``${}^T$'' also acts on time variables \nomkch{and $\TT$ is defined in \eq{eq:tt}.}
This is due to the fact that
\nomkch{
\begin{multline*}
 \bdag{A}(\tau') 
 G_0^{-1}(\tau',\tau) 
 A(\tau)  = 
 A(\tau')^T \TT 
 G_0^{-1}(\tau',\tau) 
 \TT \bdag{A}(\tau)^{T} \\ 
=
 \bdag{A}(\tau)  (\TT G_0^{-1}(\tau',\tau)^T \TT)
 A(\tau') \;.
\end{multline*}
}
Therefore, we choose $G_0$ to obey the symmetry
\beq
\label{symmg}
G_0 = (\TT  G_0^T \TT) \;.
\eeq
The same symmetry is obeyed by other
 two-point functions, such as \nomkch{the interacting Green's function}
 $G$, \nomkch{the self-energy} $\Sigma$, and their inverse. 

In principle, this redundancy
\nomkch{renders} relations 
such as \eeqref{fg0g} non invertible.
In order to avoid this, we adopt the convention that 
 functional inversion\nomkch{s} 
are carried out within 
the subspace
of two-point functions obeying the relation
 \eeqref{symmg}.
In addition, we adopt the following convention for functional derivatives
of an 
arbitrary functional $\hat \Xi$ \nomkch{with respect to} a  two-point function $X$:
\[
\FDF{\hat \Xi}{X} \to \frac12 \left(\FDF{\hat \Xi}{X}+\FDF{\hat \Xi}{\TT X^T \TT}\right)\;.
\]

\subsection{Continuum limit of the functional integral}
\label{contlim}
In principle, the expression \eq{g0tau}
should be understood \nomkch{such} that 
adjoint fields $\cdag{a}$ are evaluated at a later imaginary time 
$\tau+\delta$, whereby $\delta$ is the 
width of the discretization mesh of the interval $(0,\beta)$. The
continuum limit $\delta\to 0$ should be taken 
after having carried out the functional integration,
see, e.g.~Ref.~\onlinecite{schulman}.
 Taking this limit  at the outset amounts 
to neglecting the so-called ``contribution from infinity''.~\cite{fi.phr,be.prl} 
\nomkch{This can be achieved by} effectively replacing the normal-ordered Hamiltonian 
with a ``symmetrically ordered'' one, 
which is suitably symmetrized among possible permutation of 
creation and annihilation operators.~\cite{cc.prb}
In particular, for the noninteracting part, this amounts to replacing
the operator expression
$a^\dag a $ \nomkch{by} $  \frac12 (a^\dag a + a a^\dag) = a^\dag a +\frac12$.
Therefore, we should keep in mind that
the grand-potential $\tOmegas$ corresponds to such a symmetrized Hamiltonian.

\taglia{
\subsection{Physical dimensions}
\[ G_0^{-1}(\tau,\tau') \propto \delta(\tau-\tau') \frac{\de}{\de \tau}
\propto energy^2
\]

\[
\int d \tau' 
 G^{-1}(\tau,\tau') G(\tau',\tau'') = \delta(\tau-\tau')
\Rightarrow 
G(\tau',\tau'') \propto energy^0
\]
this is consistent with the fact that
\[
G(\tau,\tau+0^+) = density \propto energy^0
\]

\[G(\omega_n,\omega_m) \propto energy^{-1}\]

}

\bibliography{libraryea,library,references_database}

\begin{thebibliography}{52}
\bibitem[{inv()}]{inv}
\bibinfo{note}{Here, and below we assume that the relations between conjugate
  variables are invertible, at least locally, see also App.~\ref{symm}}.
\end{thebibliography}

\end{document}